**Potential Key Technologies for 6G Mobile Communications**

Yuan Yifei, Zhao Yajun, Zong Baiqing and Parolari Sergio

**Articles you may be interested in**

# Potential Key Technologies for 6G Mobile Communications

Yifei YUAN [1], Yajun ZHAO[2*], Baiqing ZONG[3], Sergio PAROLARI[4*]

[1]*China Mobile, Beijing* 100053*, China*;
[2]*ZTE Corporation, Beijing* 100029*, China*;
[3]*ZTE Corporation, Shanghai* 201203*, China;*
[4]*ZTE Corporation, Milan, Italia.*

**Abstract** The standard development of 5G wireless communication culminated between 2017 and 2019, followed by the worldwide deployment of 5G networks, which is expected to result in very high data rate for enhanced mobile broadband, support ultrareliable and low-latency services and accommodate massive number of connections. Research attention is shifting to future generation of wireless communications, for instance, beyond 5G or 6G. Unlike previous papers, which discussed the use cases, deployment scenarios, or new network architectures of 6G in depth, this paper focuses on a few potential technologies for 6G wireless communications, all of which represent certain fundamental breakthrough at the physical layer – technical hardcore of any new generation of wireless communications. Some of them, such as holographic radio, terahertz communication, large intelligent surface, and orbital angular momentum, are of revolutionary nature and many related studies are still at their scientific exploration stage. Several technical areas, such as advanced channel coding/modulation, visible light communication, and advanced duplex, while having been studied, may find more opportunities in 6G.

## 1    Introduction

Since its kickoff in March 2017, the standard development of 5G wireless communication has gone through two releases, Rel-15 and Rel-16, whose specifications are yet to be completed by the end of 2019. In Rel-15, basic functionalities, such as initial access (including beam management), channel structure (e.g., self-contained), multi-antennas (e.g., massive MIMO), and channel coding (e.g., LDPC codes and Polar codes), are specified, which can partially fulfill the performance requirements of IMT-2020. Several new technologies and scenarios, such as non-orthogonal multiple access (NOMA), ultrareliable and low-latency communication, vehicle-to-X communication, unlicensed band operation, integrated access and backhaul, terminal power saving, and positioning, are introduced to widen the use cases of 5G networks and fully support all major performance requirements of IMT-2020. Unlike previous four generations, 5G can support diverse applications, including the three main use cases, namely Gbps speed of enhanced mobile broadband (eMBB), million connection of massive machine-type communications(mMTC) and microsecond delay 99.999% level of ultra-reliable low-latency communications(uRLLC) to meet the demands of the information society in the next decade (2020–2030). The worldwide commercial deployment of 5G networks, either in non-stand-alone mode with 4G network as the anchor network or in stand-alone mode, is expected to have a significant impact on the daily life of humans, the global economy, and the culture. Like previous generations, 5G standard will continue its evolutionary path after 2020 (5G+) to further optimize the features and extend the deployment scenarios, such as non-terrestrial networks (satellite communications), unmanned aerial vehicles [1], or the operating bands, for instance, up to ~114 GHz, to invite more participation by vertical industries and emerging enterprises.

-----------------------

[*] Corresponding author (e-mail: zhao.yajun1@zte.com.cn, sergio.parolari@zte.com.cn)

Since 1982, wireless (or mobile) communication has undergone a generation change about every 10 years. Each of these 10-year cycles started with a vision (use case and deployment scenarios) and technology research at the conceptual level, followed by the standard research, specification development, prototyping of the systems, and finally, commercial network deployment. Hence, it is time to start thinking of the next generation: 6G mobile communications.

In July 2018, the International Telecommunication Union (ITU) Focus Group Technologies for Network 2030 was established to explore the system technologies for 2030 and beyond. In its concepts of 6G, new holographic media, services, network architecture, and Internet Protocol (IP) are all listed [2]. As part of the flagship program of the Academy of Finland, the 6G-Enabled Wireless Smart Society and Ecosystem (6Genesis) [3] was founded in 2018, which is focused on the study of wireless technology and the standard development of 6G communication. Its research areas span over reliable, near-instant, unlimited wireless connectivity; distributed computing and artificial intelligence; and materials and antennas for future circuits and devices. The United States also expressed its 6G ambition through an announcement by a Federal Communications Commission official at the 2018 Mobile World Congress [4]. In China, according to an interview with the Minister of Industry and Information Technology in March 2018, study on 6G has already begun in the country [5]. Elsewhere, the European Union, Japan, South Korea, Russia, and other countries have also started to carry out relevant work.

Use cases, deployment scenarios, and performance requirements of 6G were envisioned in several papers [6][7][8][9]. New network architectures were also discussed [10][11]. Regarding the potential technologies, especially at the physical layer, previous generations of mobile communications were normally hallmarked by multiple-access schemes, such as frequency-division multiple access (FDMA), time-division multiple access (TDMA), code-division multiple access (CDMA), and orthogonal frequency-division multiple access (OFDMA). This highlights the importance of technology advancements, which are not only related to the air interface designs but also the results of various breakthroughs in electronic/photonic materials, microelectronic fabrication, and device manufacturing. For instance, circuit digitization makes the shift-keying signals and channel coding possible, thus significantly increasing the voice capacity in TDMA-based 2G systems (e.g., GSM) compared to FDMA-based 1G systems. The migration from digital signal processing (DSP) to application-specific integrated circuit (ASIC) drastically elevates the processing power and density of the baseband in the base stations, which is crucial for the capacity gain of 4G systems compared to 3G systems. The highly integrated circuits in the baseband and radio frequency and optical fiber domain make active antennas feasible in engineering, and then turn the academic-world massive MIMO into a reality in 5G.

Given the rather early stage of 6G research, openness in technology should be encouraged, similar to use cases, deployment scenarios, and performance requirements. The choice of technology also reflects the investment in some strategic areas of a country, which is especially the case in 5G. Considering the abovementioned instances, in this paper, we focus on potential physical layer technologies for 6G, which include holographic radio, terahertz communication, large intelligent surface (LIS), orbital angular momentum, advanced channel coding/modulation, visible light communication, and advanced duplex. The maturity levels of these technologies vary, some of which are still in their scientific exploration stage. Note that artificial intelligence (AI) or machine learning (ML) will definitely play an important role in 6G and can be used widely in many technical fields. Hence, it will be discussed in conjunction to the abovementioned potential technologies, rather than a separate section dedicated to AI.

The rest of this paper is organized as follows. 6G concepts are discussed briefly in Section 2. Section 3 is devoted to the four revolutionary technologies of exploratory nature: holographic radio, terahertz communication, large intelligent surface, and orbital angular momentum. In Section 4, three more matured technologies are discussed: advanced channel coding/modulation, visible light communication, and advanced duplex. The summary is provided in Section 5.

## 2  6G concepts

- **6G vision**

The goal of 6G is to meet the demands of the information society 10 years from now, e.g., ~2030, which would significantly go beyond what 5G can offer. 6G vision can be summarized into four key aspects, "intelligent connectivity," "deep connectivity," "holographic connectivity," and "ubiquitous connectivity," which constitute the overall vision of 6G, which is "Wherever you think, everything follows your heart."

"Intelligent connectivity" refers to the inherent intelligence of communication systems: intelligence of network elements and network architecture, intelligence of connected objects (terminal devices), and information support of intelligent services. 6G networks will face many challenges, such as super complex and immense networks, myriad types of terminals and network devices, and extremely complex and diverse business types. "Intelligent connectivity" will meet two requirements simultaneously: 1) each of the related connected devices in the network itself is intelligent and the related services are intelligent, and 2) the complex network itself needs intelligent management. "Intelligent connectivity" will be the fundamental characteristic for supporting the other three major characteristics of 6G network: "deep connectivity," "holographic connectivity," and "ubiquitous connectivity."

- Requirements and KPIs

To realize the vision of the 6G network and to meet the demand of future communications, the following key requirements and challenges need to be considered, particularly compared to some KPIs of 5G:

- Peak rate: terabit era, ~10 terabits per second, which is ~10 times higher than 1 terabits per second for 5G networks with system bandwidth of ~hundred MHz
- Universal connection with low delay, reliability, and high rate
- Higher energy efficiency, compared to no definite requirements in 5G either at the network side, or the terminal side if for eMBB and URLLC
- Connection everywhere and anytime, as opposed to a million devices per square kilometer for 5G which may not be very challenging, depending the assumptions of traffic models, system bandwidths, etc
- Ubiquitous intelligence, vs. no such requirement in 5G
- Native security (trust)
- Versatility: to accommodate various networks in dynamic and organic manner
- Convergence of communication, computing, sensing, and control
- Nontechnical challenges: industry barriers, policy and regulation, and consumer habits

- Enabling technologies

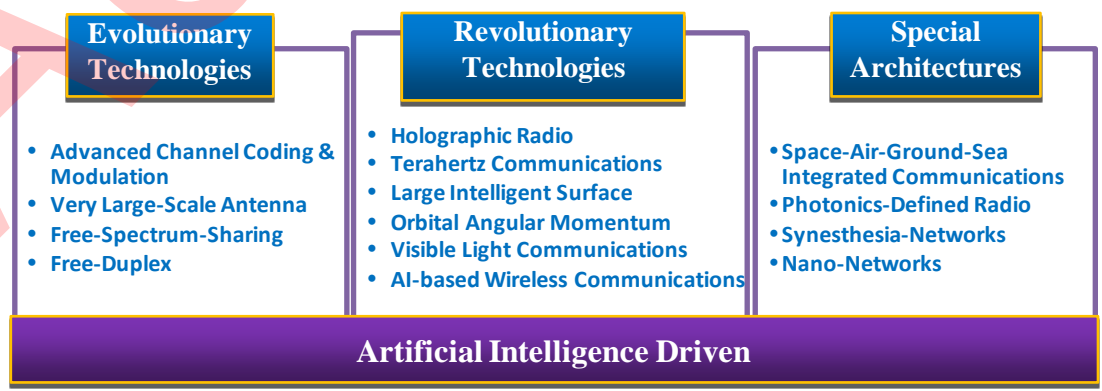

Figure 1 Potential key technologies of 6G

Currently, the concept of 6G is still in the early stage of discussion, and the views expressed by different countries are quite diverse. To achieve the abovementioned 6G vision, and by considering the development status and trends of related technologies, we believe that the potential key technical features of 6G would include the aspects shown in Figure 1. Among them, for instance, Terahertz communications, visible light communications, very large-scale antenna, advanced channel coding would be crucial for achieving the peak rate of ~10 terabits per second and the extreme low latency. Advanced channel coding & modulation and space-air-ground-sea integrated communications can help to truly fulfill the massive connection (everywhere and anytime). Holographic radio and large intelligent surface are promising in significantly improving the energy efficiency and reducing the hardware cost of 6G networks.

In previous generations of mobile communications, it would sometimes be difficult to clearly categorize whether a technology should belong to the a-th or b-th generation. Similarly, some technologies may be envisioned as the evolution of 5G, since the level of maturity is higher and the related study has been conducted for many years, e.g., the problems, challenges, and general approaches are well known. Depending on the demand of future releases of 5G, there is a chance that some of these technologies may ultimately be part of 5G.

On the other hand, several technologies are still in the exploratory stage—strictly speaking, some of them are still science, not technology. Some may highly depend on the advancement of other fields, such as material science, physics, chemistry, and semiconductor manufacturing. However, they may not reach maturity before 2030.

Hence, in the following, we discuss these two categories, revolutionary technologies and evolutionary technologies, in separate sections.

## 3 Potential revolutionary technologies

The technologies to be discussed in this section are revolutionary in the sense that they would fundamentally change the physical layer of mobile communication systems compared to 5G. Many aspects are still at the stage of scientific exploration. Yet they indeed represent the level of science and technology development of a country in cutting-edge strategic areas.

### 3.1 Holographic radio

Today's 5G follows the technological path of previous mobile communication systems, and its performance is hitting a wall due to the lack of fundamental breakthrough in the physical layer in recent years. Therefore, new theory and paradigms and innovative breakthrough technologies are needed for 6G. There are some new concepts, such as photonics-defined radios and holographic radios, on the horizon [7][11]. Among them, holographic radios are likely to significantly improve the efficiency of spatial multiplexing and achieve holographic imaging level and ultra-high-density and pixelated ultra-high-resolution spatial multiplexing [7][12].

It is known that holography records the electromagnetic field in space based on the interference principle of electromagnetic waves. The target electromagnetic field is reconstructed by the information recorded by the interference of reference and signal waves. The core of holography is that the reference wave should be strictly coherent as a reference, and the holographic recording sensor should be able to record the continuous wavefront phase of the signal wave so as to accurately record the high-resolution holographic electromagnetic field [13]. Because radio frequency (RF) and light waves are both electromagnetic waves, holographic radios are very similar to optical holography. For holographic radios, an antenna is the usual holographic recording sensor, so a continuous aperture antenna array is needed to receive and measure the continuous wavefront phase of the signal wave. To implement a continuous-aperture antenna array, one method is to use a conventional discretely spaced antenna array, but the number of elements is close to infinite, that is, $N \rightarrow \infty$. This is obviously unrealistic and will be a disaster for the system's size, weight, and power (SWaP). Another solution is to integrate several antenna elements into a compact space in the form of a spatially continuous electromagnetic aperture, a so-called meta-surface. However, this method is limited to passive reflective meta-surfaces, because for a continuous-aperture active antenna array, the RF feed network is simply impossible to achieve due to the ultra-dense elements. A pure passive meta-surface cannot be used as stand-alone to build a complete radio access network, so it can only be an auxiliary and supplementary unit for radio access networks.

To achieve a continuous-aperture active antenna array, a promising method is to use an ultra-wideband tightly coupled antenna array based on a current sheet. The uni-traveling-carrier photodiodes (UTC-PDs) are bonded to the antenna elements through flip chip technology and form coupling between the antenna elements [14]. In addition, the patch elements are directly integrated to the electro-optic modulator [15]. The current output by the UTC-PD directly drives the antenna elements, so the entire active antenna array has a very large bandwidth (~40 GHz). Moreover, this continuous-aperture active antenna array does not require an ultra-dense RF feed network at all, resulting in not only feasible implementation but also obvious SWaP advantages. Holographic radio technology transmits and receives radio signals through a spatially continuous aperture formed by advanced UTC-PD antenna array technology, transforming the limited beam space of a traditional antenna array into a plane wave of nearly infinite beam space, that is, to achieve a nearly infinite, continuous multiplexing space or beam space, compared to the traditional massive MIMO, which is a discrete aperture and limited beam space.

Another difference of holographic radios from traditional massive MIMO is that they use Fresnel-Fraunhofer interference, diffraction, and spatial correlation models instead of traditional Rayleigh scattering propagation models to model and perform channel estimation on holographic channels. An accurate computation of communication performance requires detailed electromagnetic numerical computations, that is, algorithms and tools related to computational electromagnetics and computational holography. Generally, the spatial correlation propagation model is described based on the Fresnel-Kirchhoff integral. Figure 2 shows the comparisons between holographic radio and massive MIMO.

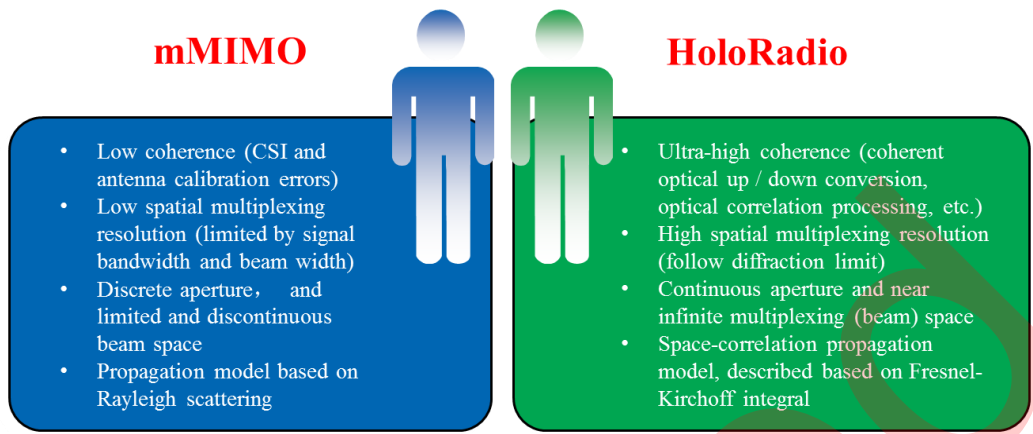

**Figure 2** Comparison between holographic radio and massive MIMO

Holographic radio can not only be used to achieve the RF holography, spatial spectral holography, and spatial wave field synthesis but can also precisely regulate the electromagnetic field of the entire physical space in a fully closed loop. It can also significantly improve the spectral efficiency and network capacity and even realize RF convergence of imaging, positioning, and wireless communications [7][16]. However, an extreme wideband spectrum and holographic RF generation and sensing will generate considerable data. Although these massive data are generally useful for ML to efficiently train and learn, a low-latency, highly reliable, and scalable artificial intelligence architecture is needed to process the massive data of holographic radio. Therefore, for a 6G system that combines the full spectrum, AI, and RF holography, if traditional electronic signal processing and computing are to be used, its SWaP and latency will be huge challenges. To meet the challenges of 6G energy efficiency, latency, and flexibility, a hierarchical heterogeneous optoelectronic computing and signal processing architecture is inevitable [7][17][18]. Holographic radios achieve ultra-high coherence and high parallelism of signals through the coherent optical up-conversion of the microwave photonic antenna array, and this ultra-high coherence and high parallelism facilitates the signal to be processed directly in the optical domain. Considering that optical computing is more suitable for linear computations [19], more than 90% of holographic radio signal processing involves real-time linear computations in the optical domain, which is the key to achieving 6G air interfaces with high energy efficiency and low latency. Figure 3 shows the architecture of an all-photonic holographic radio system.

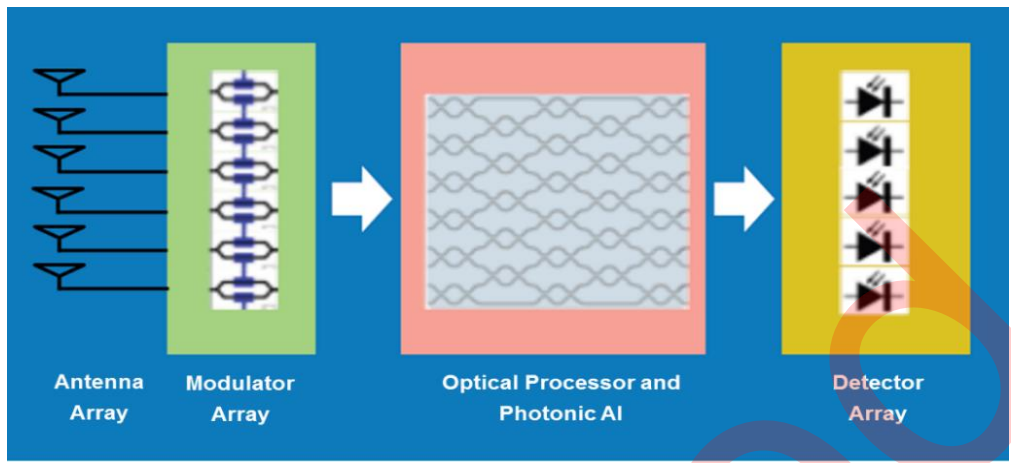

**Figure 3  Architecture of an all-photonic holographic radio system**

### 3.2  Terahertz communications

The terahertz frequency band ranges from 0.1 to 10 THz, which is the last span of the radio spectrum and generally considered as a terahertz gap. The terahertz band is envisioned to provide with up to Tbps data speed to satisfy extremely high throughput, low latency, and completely new application scenarios for 6G [20][21] which may not be possible in millimeter-wave bands where the system bandwidth can rarely exceed 1 GHz. The first project in IEEE 802 toward 100 Gbps, IEEE 802.15d, was approved in March 2014, although there is no commercial plan based on this standard. The unique characteristics of the terahertz band, such as high path loss, scattering, and reflection, pose many new challenges that need to be addressed before achieving the Tbps (terabit(s)) links.

- **Characteristics of terahertz bands**

The terahertz spectrum has some unique advantages for wireless communications, but there are also many challenges. The benefits include huge bandwidths (>50 GHz) available to support the Tbps links, higher frequency (short wavelength) with spatial resolution, and short pulse (picoseconds) with high resolution in time domain, which can be used for super-resolution sensing and high-precision positioning. The challenges of THz communications lie in the wireless propagation characteristics of the THz spectrum. (1) Severe path loss even for free-space propagation, e.g., ~100 dB at 300 GHz at a distance of 10 m. (2) Excessive attenuation due to resonance of molecules in the air [22]. Note that there are several atmospheric windows, e.g., 140, 220, and 340 GHz [23][24], where the attenuation due to molecule resonance is relatively benign, e.g., only ~2 dB/km, negligible compared to the free-space attenuation. In addition, when the frequency exceeds 1 THz, the radio wave undergoes a significant absorption by water vapor and oxygen molecules in the atmosphere and can be attenuated ten times at 1-m propagation distance [25]. (3) Sensitivity to shadows and blocking due to the weak diffraction effect at such short wavelength. For example, the signal attenuation of a brick can be as high as 40–80 dB, and the human body can cause 20–35 dB signal attenuation. (4) Less sensitivity to humidity/rainfall, e.g., attenuation becomes relatively flat above 100 GHz. 5) Superfast channel fluctuation and intermittent connection, e.g., the coherence time of the terahertz band is very short, and the Doppler frequency is very high.

Besides the inherent characteristics of the THz spectrum mentioned above, other challenges in engineering implementation need to be considered. For example, ultra-high processing power may be needed to handle the extremely wide bandwidth and very large-scale antenna. Consequently, it is necessary to design a superfast-speed broadband processing chip, which leads to extremely high power consumption. This is because power consumption is generally proportional to the sampling rate and the broadband terahertz system (A/D) conversion. The THz power amplifiers (PAs) are another significant obstacle, as the state-of-the-art PA efficiency is still quite low at such high frequency, making it harder to fit the PAs into practical base stations and user equipment (UE).

Nevertheless, the manufacturing of THz components is getting more mature, and some commercial products become available, which can emit power of 0–10 dBm at 300 GHz. It is expected that higher output power can be achieved in the near future. Moreover, simple modulation schemes (e.g., BPSK and QPSK) are enough for high data rates (>100 GBit/s), given the huge bandwidths of THz bands, at least in the first stage. Therefore, THz communications have an opportunity to be applied in the 6G era, and we can look forward to further improving the performance with some key technology breakthroughs.

- **Targeted application scenarios**

The terahertz communication scenarios can be mainly classified into macro-, micro-, and nanoscale networks. Macro-scale networks are primarily used for the applications in which the transmission range is from 10 m to few kilometers. Micro-scale networks are typically used in applications with limited transmission range (less than few meters, e.g., $\leqq 10$ m). Nanoscale networks are more suitable for communications within a range of below 1 m or cm.

Macro-scale networks are often used for outdoor scenarios, and the typical applications include vehicle-to-vehicle connection and backhaul/fronthaul connection. These applications would need a wider coverage (e.g., 10 m to a few kilometers) and high throughput (up to 1 Tbps) with low latency (e.g., <1 ms).

Regarding micro-scale networks, they can be further categorized into outdoor and indoor scenarios. For indoor scenarios, applications requiring mobility, as well as the applications with fixed point-to-point or multi-point connections, such as indoor small cells, wireless personal area network, wireless connections in data centers, and near-field communications, such as kiosk downloading, can be supported. For outdoor scenarios, the applications can include vehicular, small cells, and backhaul connection, while for the indoor environment, it is different because of the reflections and scattering phenomenon with path and absorption loss. Therefore, these indoor and outdoor scenarios require different propagation models to represent different obstacles, scattering, and atmospheric losses.

The nanoscale network is a novel network topology suitable for extremely short wavelengths. In a nanoscale network, communication is usually for a distance within 1 m or cm (e.g., inter-miniature-device links, on-chip and chip-to-chip links, in-body communications). The main challenges include the novel transceiver design for nanoscale devices, channel models for the new propagation environments of the nanoscale networks, physical layer solutions including channel coding and modulation schemes, and communication protocols.

Besides the above three terrestrial scenarios, outer space communications are also important scenarios of terahertz communications, which have been applied to the field of space science for many years. In outer space, within terahertz bands, the relatively transparent atmospheric windows are ~350, 450, 620, 735, and 870 μm, where the transmission would experience little moisture-induced absorption, e.g., long-distance communication becomes feasible.

- **Key technology aspects**

Although the technology for terahertz communication is evolving rapidly in the areas of transceiver architectures, materials, antenna design, propagation measurement, channel modeling, and physical layer techniques, there are still many challenges to be addressed before Tbps links become practical. The following are the key enabling technologies/studies for THz communication.

- Propagation measurement and channel modeling
    - Different characters of the scattering and reflection due to the very short wavelength
    - Molecular and water vapor absorption, atmospheric windows
    - Indoor, outdoor, intra-device, in-body, and outer space
- THz signal generation and detection (transmitter and receiver designs)
    - Electronics-based approach and photonics-based approach, considering the transmitting power and power efficiency, complexity, cost, and size
    - Semiconductor technology and meta-materials, e.g., graphene-based electronics [26]
- Antenna technologies
    - Nano- and meta-materials for plasmatic antenna arrays, e.g., graphene-based antennas [27]
    - Synchronization mechanism, such as high-speed and high-precision acquisition and tracking mechanism and synchronization mechanism for antenna arrays with hundreds or thousands of antenna elements
- Ultra-massive MIMO for THz band, leveraging LIS-assisted smart radio environment

- Beamforming to overcome the severe path loss
- Spatial multiplexing for high data rate in short-range/nano-communication
- Leveraging LIS-assisted smart radio environment to overcome the blockage issues caused by the directional beam (for details of LISs, refer to Section 3.3)

- High-speed baseband signal processing technology
    - Research on high-speed baseband signal processing technology with low complexity and low power consumption and integrated circuit design to develop terahertz high-speed communication baseband platform

- Baseband design
    - Waveform, e.g., single carrier for lower cubic metric (CM)
    - Modulation and channel coding
    - Multiple access (e.g., OMA and NOMA)

Among the abovementioned aspects, transmitter design for THz is the most critical. There are two types of devices that can perform conversion to terahertz signals: electronics-based devices and photonics-based devices. A key aspect of terahertz communication is carrier frequency generation, where two approaches are being studied: photonics based and electronics based.

Photonics-based techniques offer the unique advantage of ensuring a high modulation order obtained from the optical-to-THz conversion with photo-mixing, high-speed amplitude, and/or phase coding introduced from optical coherent network technologies. A unique feature of using photonics devices is the possibility to easily address multi-carrier transmission by adding optical laser fibers to the optical driving signals. THz communication systems can deliver high-rate data wirelessly, using photonics-based approaches. Due to the intrinsic high propagation loss at higher carrier frequencies and the low power generated at these frequencies by photonic sources, so far, the transmission distance is usually within a very short range (e.g., ~10 m at 409 GHz [28]). To tackle the power limitation of photonics devices, the future photonics-based THz systems may be based on the combination of PAs associated with photo-mixers. Better performance would require a monolithic association of the two devices, which has not been achieved in the THz range. In the scenario of photonics-based THz transmitters for spectrally efficient data links, the optical feed/source deserves some special attention where the spectral content of the THz signal is directly fed to the two optical laser lines that drive the photo-mixing devices, such as photodiodes. The overall performance would depend on the optical feed, which features low jitter and narrow linewidth in the optical domain.

The electronics-based approaches are now feasible for most of the 100–150 GHz-band wireless links. These approaches attract significant interest due to their room-temperature operation, compact size, and readiness for chip integration. In these approaches, frequency multiplication is the most commonly used method for generating terahertz frequency signals. So far, at frequencies above 100 GHz, GaAs and InP ICs have been key materials in all-electronic THz communication research. This is mainly due to the high cutoff and maximum frequencies of transistors. However, other technologies are also good candidates for practical THz communications and mass production-compatible chipsets are now being pursued by many companies and institutes. For example, Si-IC technologies have started to show their THz potential in the last 2–3 years. The major limitations of Tx and Rx of electronics-based approaches for the THz communication system include the following:

- The nonlinear behavior of multiplier chains, which limits the modulation to be amplitude only.

- The limited modulation choices if the modulation is conducted with a sub-harmonic mixer at the source output.

- The relatively high impedance (~k$\Omega$) of Schottky barrier diode-based direct detectors, which limits the achievable bandwidths, even if some trans-impedance amplifier integration can partially overcome this limitation.

Among these approaches, the highest data rates have been obtained using photonic devices as transmitters, combined with cutting-edge III/V THz electronic devices as receivers.

Besides the abovementioned pure photonics- and electronics-based approaches, combined mechanisms were also studied. The combination of photonics and electronics active devices, a truly hybrid microwave photonics approach, can benefit from the intrinsic advantages of each of the two technologies, such as tunability, fiber compatibility, and power handling capability.

A suitable scenario can be selected based on the required data rate, propagation distance, and detection sensitivity, which can balance the main key factors, such as bandwidth, tunability, stability, and fiber compatibility, from photonics and power-handling capability from the electronics. For example, for long-distance communication scenarios (e.g., >100 m), electronics-based approaches with higher Tx power may be suitable, while for hotspot/indoor scenarios and nano-communications (e.g., less than 10 m), photonics-based approaches with lower Tx power may fit better.

In addition, ITU has formally designated 0.12–0.2 THz for wireless communications [29] and has just identified applications in the frequency range of 275–450 GHz in ITU-R WRC-19. However, the detailed rules and regulations for this new spectrum are neither well developed nor with a globally unified understanding. This requires the joint efforts of ITU and WRC to actively promote the global consensus.

### 3.3 Large Intelligent Surface

A drastic increase in the spectral efficiency, one of the key requirements of 6G, can be achieved by leveraging the combined benefits of a high spatial multiplexing gain from massive MIMO and high bandwidth of the THz band. However, several radio frequency (RF) chains operating in high-frequency bands will lead to overcomplicated signal processing, extremely high power consumption, and prohibitive hardware cost. An LIS is a promising energy-efficient and cost-effective solution for tackling the above challenges. It is envisioned that an initial leap from traditional massive MIMO toward LISs can provide LIS-assisted smart radio environments and generate a completely new network paradigm for 6G networks [30][31], as seen in Figure 4.

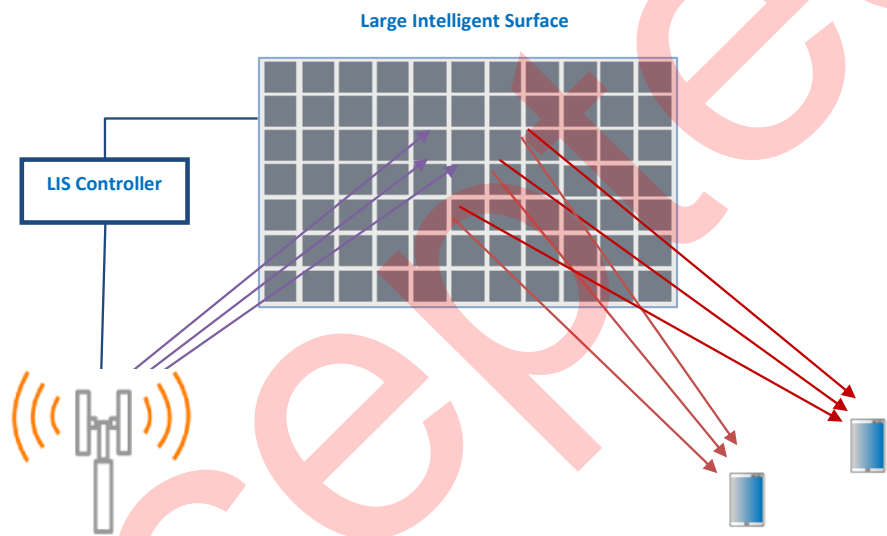

**Figure 4 LIS-assisted smart radio environments**

- **Comparison with Traditional Technologies**

An LIS is an artificial surface made of electromagnetic materials, which can change the propagation of incoming and outgoing radio waves. It is significantly different from other traditional technologies, such as massive MIMO and amplify-and-forward relay (AF relay).

An LIS can be seen as an extension of the massive MIMO, but goes well beyond the traditional antenna array concept. LISs are different from the massive MIMO due to their different array architectures (passive devices versus active devices) and operation mechanisms (reflection versus transmission). LISs can achieve unprecedented performance gains of massive MIMO while consuming much lower energy, due to the passive nature of the elements [32]. In [33], the information delivery capability of an LIS was analyzed, where the authors proved that the capacity per unit surface area of an LIS linearly grows with the average transmit power, rather than a logarithmic relationship in the case of large-scale MIMO deployment.

In some sense, LISs may resemble the classic AF relays; however, there is a big difference between them [34]. Relays are formed of active elements (e.g., PAs), which reduce the overall data rate of the two-hop communication if in half-duplex mode or are subjected to severe self-interference if in full-duplex mode. In contrast, LISs acting as reconfigurable reflectors with passive elements have the advantage of low power consumption, since they only reflect the signals passively without involving active RF chains. In addition, they do not have issues of self-interference and noise amplification effects, which are the two most important shortcomings of AF relay.

As the key components of LISs, passive reflectors have been used in radar systems for many years. However, in most radar systems, the phase shift of passive elements cannot be changed once they are manufactured, which prevents them from being used for the wireless channels that are often time varying. In fact, for 6G, each of LIS's passive reflectors should be able to independently tune the phase shift of the signal incident on it, thereby creating a favorable wireless transmission channel. By properly tuning the phase shift via an LIS controller, the reflected signal can be superimposed constructively at the target receiver to boost the received signal power while suppressing the reflected signal at the un-target receiver to reduce co-channel interference.

- **Advantages**

LISs are made of low-cost passive elements that do not require any active power source for transmission. Their circuitry and embedded sensors can be powered with energy-harvesting modules as well. They primarily rely on the programmability and re-configurability of the intelligent meta-surfaces, as well as their capability of shaping the radio waves impinged on them in a controlled manner. They can operate in the full-duplex mode with little self-interference without increasing the noise level.

The ability of controlling the response to each LIS and choosing its location via a software-programmable interface allows the optimization of wireless networks without the constraint of the underlying physics of wireless propagation and the meta-materials. This enables the seamless integration of LISs into software networks. Further information about the programmability via software and the integration of LISs into software networks can be found in [35]. Compared to other solutions, the use of LISs has enormous economic impact, e.g., they reduce the waste of hardware resources and offer more accurate control of the radio waves and high scalability of the deployment.

By intelligently tuning the phase shifts induced by the elements of LISs, the LISs can achieve multiple objectives, such as overcoming unfavorable propagation conditions and enriching the channel by introducing more multi-paths and increasing the coverage area, while consuming very little energy. For example, the high-frequency systems, including millimeter-wave and beyond 100 GHz communications (i.e., THz bands), can take advantage of an LIS as a source of controllable reflectors that can mitigate NLOS propagation conditions. It can act as reconfigurable reflectors to establish strong NLOS links where LOS is not available or is just not sufficiently strong to achieve a reliable connection. By doing so, the coverage and rate of the wireless systems can be improved.

LISs have other advantages as well, such as low profile, light weight, and conformal geometry, which make it easier to be installed on man-made structures (e.g., walls, ceilings, and roads) and natural objects, thus providing a high degree of flexibility and superior compatibility for practical deployment, to enable smart radio environments.

All the above advantages render LISs an appealing solution for performance advancement in 6G networks, especially for indoor applications with a high density of users in some scenarios such as stadiums, shopping malls, exhibition centers, and airports.

- **Key enable technologies**

There are many challenges that need to be addressed before a smart wireless environment can be achieved:

- Meta-surface design to study on a more flexible and powerful programmable LIS.

- LIS-based wireless network paradigm and deployment.

- LIS channel sensing/feedback and control, such as channel state information (CSI) measurement and analog beamforming using LISs.

- Combination with massive MIMO, terahertz, and other related technologies, e.g., LIS-assisted massive MIMO.

From the perspective of network deployment, the LIS-assisted network introduces a new paradigm. How to optimize the deployment of the LIS-assisted network with passive reflectors and meta-surfaces is a new challenge. One solution is the AI-powered operation of a reconfigurable LIS. A fundamental analysis is needed to understand the performance of LISs and smart surfaces, in terms of rate, latency, reliability, and coverage. Another important research direction is to investigate the potential of LIS-based reflective surfaces in terms of enhancing the range and coverage of tiny cells and dynamic modification of the propagation environment.

To materialize the concept of a smart radio environment, a key problem to be addressed is that in order to configure and optimize the environment according to the network conditions, the amount of sensing data collected on the meta-surface and the amount of feedback for the overall network controller can be huge. Effective solutions need to be developed to reduce the amount of sensing data required for network optimization, e.g., only enough information is to be provided to network controllers with low overhead and high energy efficiency. In [36], a novel approach by leveraging tools from compressed sensing and deep learning is proposed to solve the problems with low-complexity hardware architectures and low overhead for training.

### 3.4 Orbital angular momentum communication

Orbital angular momentum (OAM), known as "vortex electromagnetic wave," has attracted much research attention [37]. Unlike the traditional plane electromagnetic (PE) wave-based signals, radio vortex signals have the phase rotation factor $exp(-jl\varphi)$. The main advantages of OAM are that the electromagnetic wave characteristics associated with beam vorticity and phase singularity can have an unlimited number of eigen states (i.e., OAM modes), which are orthogonal to each other, and therefore, multiple channels are allowed to increase the transmission capacity and spectral efficiency in principle [38]. OAM opens up a new dimension for electromagnetic wave multiplexing transmission, which bridges a new way to significantly increase the spectrum efficiency and is expected to be used in future wireless communication networks (6G).

Originally, the study of OAM was limited to the field of optical communication. Only in the last decade, low-frequency electromagnetic waves, such as microwave, millimeter wave, and terahertz band, have been considered for OAM. In [37], the application of photon orbital angular momentum to low frequency was proposed. Through simulation, it was proved that the phased array antenna can be used to generate eddy electromagnetic waves similar to Laguerre-Gauss eddy beams. In recent years, some experiments have proved the feasibility of wireless communication based on OAM [39][40]. In [41], the author proved experimentally and theoretically that the combination of MIMO-based spatial multiplexing and OAM-based multiplexing can improve the spectral efficiency of wireless communication. In this paper, the author proposes a method of rotating OAM electromagnetic wavelength distance transmission and successfully carries out the first 27.5 km transmission experiment in the world [42].

Due to the introduction of a new resource dimension, some traditional wireless communication concepts may need to be updated, and related mechanisms can be redesigned accordingly. There are three basic advantages of OAM-based wireless communications: (1) high spectral efficiency via new domain of OAM modes; (2) supporting more users to access the system since OAM provides a novel multiple access method, i.e., mode division multiple access, without consuming more frequency and time resources; and (3) high reliability for anti-jamming because OAM can not only be used within the narrow band but also be jointly used with frequency hopping over a wide band to improve the ability of anti-jamming for wireless communications.

- **Comparison with conventional MIMO**

In [43], it is claimed that radio communication over the sub-channels of OAM states is only a subset of spatial dimensions offered by MIMO. In [44], it was verified that, with the constraints of the receiver size, an OAM-based MIMO radio system is equivalent to conventional MIMO systems from the perspective of spatial channel multiplexing. So far, no one has proved that a practical OAM for multiplexing has higher link capacity than the conventional single-user MIMO. However, unlike the plane waves used by MIMO, the wave vector of OAM is along the azimuthal direction. Through this particular wave vector, the OAM-based MIMO system can be seen as a conventional MIMO system to increase the spatial dimension. Therefore, this system has a great potential in a massive MIMO method, which requires several antennas and limited spatial spacing between antennas. OAM-based MIMO is also very suitable for the communications in an open area or long-distance communications. For example, it is found that such an OAM-based MIMO system can increase the communication distance for the line-of-sight (LoS) MIMO channel if the OAM states as the elements of the transmitting ULA are sorted in an ordered sequence.

- **Technological challenges and future research**

The feasibility of wireless communication based on OAM was validated; however, there are still many unresolved research issues. To further promote the application of OAM in future wireless communications, there are three aspects that require a comprehensive and thorough study.

Multiplexing and de-multiplexing of OAM radio waves. The multiplexing and de-multiplexing of OAM radio waves are large bottlenecks for OAM-based wireless communication. The OAM optical beams can be deftly multiplexed and de-multiplexed. However, when it comes to the radio frequency regime, as the wavelength is much longer than the that of the optic waves, it is difficult to manipulate the OAM radio beams, such as beam combining and splitting, meaning that the coaxial transmission cannot be easily ensured. Thus, specially designed devices are needed and may bring great insertion loss in the link, which would ultimately reduce the efficiency [45].

Radio vortex signal transmission. Three typical problems need to be considered for radio vortex signal transmission. (a) Transmitter–receiver alignment. The transmitter and receiver must be aligned with each other to separate signals with different OAM modes. In non-aligned scenarios, the phase turbulence adaptive estimation algorithm needs to be implemented at the receiver. (b) Fading. There exists fading, such as atmospheric turbulence, rain, and fog [46], which can disturb the wavefront phases at the receiver. (c) Convergence. An OAM beam becomes increasingly divergent as the order of the OAM mode increases; it severely reduces the transmission distance and decreases the spectrum efficiency of OAM-based wireless communications. It is critical to make OAM beams convergent so that all OAM modes, including both high- and low-order OAM modes, can be efficiently used.

Radio vortex signal reception. At the receiver, the phase detector tries to distinguish the order of different OAM modes. How to effectively separate and detect the information modulated on the eddy electromagnetic wave is one of the key challenges. The radio vortex signal reception schemes for SPP antenna, UCA antenna, as well as other schemes, i.e., the phase gradient method (PGM), are developed to identify the OAM modes.

There are some other issues that need to be considered, including limited number of available OAM modes and joint OAM mode and frequency/time partition.

Although the OAM beam is vorticose hollow and divergent, the divergence of OAM beams reduces significantly as the frequency increases. Thus, OAM may find more use in networks with high frequency, such as terahertz bands.

## 4 Potential technologies with more maturity

The technologies introduced in this chapter reach certain maturity compared to those discussed in Chapter 3. Nevertheless, significant changes in the physical layer are still needed to support or accommodate these technologies.

## 4.1 Advanced channel coding and modulation

As fundamental physical layer technologies, channel coding and modulation provide efficient ways for a radio link to operate near its channel capacity, while making the signal waveforms friendly to RF and baseband processing at transmitters and/or receivers. Since the advent of 2G, almost each generation is marked with or dominated by a new channel coding scheme, for instance, convolutional codes in 2G, Turbo codes in 3G, enhanced Turbo codes in 4G, LDPC codes, and Polar code for 5G. In general, channel coding and modulation encompass many specific fields that may touch waveform or even multiple accesses. The mathematic tools are also very diverse, for example, combinatory algebra and number theory for channel coding, linear algebra for waveform, and detection theory for modulation.

- **Channel coding**

Traditionally, channel coding design assumes a single link connection whose channel can be binary symmetric, binary erasure, or additive white Gaussian noise. The performance target is the Shannon limit of single-link channels. Such design makes sense in many mobile systems since 2G, as the primary multiple access schemes are orthogonal where different users (at least served by the same base station) occupy different time-frequency resources, or in non-overlapped spatial domains. However, to increase the system capacity and serve more connections, non-orthogonal multiple access can be used as a complement to orthogonal multiple access. This opens an entirely new area: multiuser-oriented channel coding, which can be considered as a significantly enhanced version of interleave-division multiple access (IDMA). It is new also in the sense that the channel capacity of the multiuser channel is not entirely known or proved, especially for the uplink with near-far effect and independent channel fading. In theory, any channel code can be considered with the purpose to increase the capacity of multiuser channels, as long as the code itself can provide enough room for optimization, e.g., from single user to multiuser, as illustrated in Figure 5. Recently, multiuser LDPC codes were proposed for uplink non-orthogonal transmission [47]. The reasons of considering LDPC codes are as follows. First, quasi-cyclic binary LDPC is already specified in 5G as the channel coding scheme for data channels. Its performance superiority and low decoding complexity are well demonstrated for medium and long code blocks, compared to other major channel coding schemes. For a multiuser channel, the receiver complexity is generally higher than that of a single-user channel. Thus, codes with low decoding complexity are very attractive; second, LDPC has many design flexibilities, particularly, the proto-matrix for parity check, the matrix lifting, and the shifts. With proper choices of these parameters, binary LDPC has great potential to offer performance benefit for a multiuser channel.

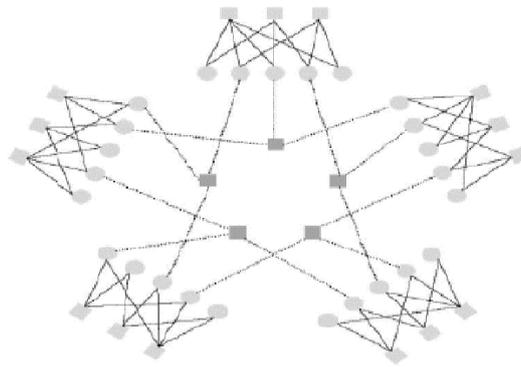

**Figure 5 Bipartite structure for multiuser LDPC**

In previous generations of mobile communications, most channel coding schemes at the physical layer operated in the binary domain. To increase the robustness of the channel codes in fading channels and to operate in very high signal-to-noise ratio (SNR) scenarios, non-binary (also called multivariant) codes can be considered. Two types of multivariant codes are already prominent. The first one is multivariant LDPC, proposed by Davey and MacKay [48]. Such codes are defined in the Galois field (q). The basic design of multivariant LDPC resembles that of binary LDPC; for example, the parity matrix can be randomly constructed or follow certain defined patterns such as quasi-cyclic, with belief propagation or its variants as the basic decoding algorithm. Its design complexity and decoding complexity are generally higher than those of binary LDPC. However, its ability of removing the "short loop" in the partite graph of the parity matrix allows the code to effectively combat the burst errors. The second type of such codes is the lattice code. A very promising variant of the lattice code is the low-density lattice code [49], which can also be represented as the Tanner graph and parity check matrix. Therefore, the belief propagation algorithm can be used, whose decoding complexity grows only linearly with the length of the code block.

Most of the channel codes are designed with limited choices of the coding rate, where the performance of the codes can be optimized for specific code rates. While it can support many coding rates, quasi-cyclic LDPC for 5G is still not rate-less. Spinal codes [50] are true rate-less codes that can provide consistently good performance over a wide range of code rates. The attractiveness of the spinal codes also lies in the near-Shannon capacity performance when the code block is short, as well as the superior performance when SNR is very high.

Simple enhancements of the 5G LDPC codes or Polar codes, or 4G turbo codes or convolutional codes, can also be considered, for example, to improve the performance when the code block length is short for the traffic channels, to make the performance more robust various environments.

- **Modulation and spreading**

It is well known that the distribution of QAM constellation points is far from Gaussian. Capacity wise, QAM is not a perfect choice even though its signal generation and demodulation are simpler than many other modulations. Amplitude phase shift keying (APSK) has been used in various broadcasting networks and satellite communications. APSK is very robust to the nonlinearity of the PA. APSK can also tolerate a higher level of phase noise, compared to other modulation schemes. Therefore, APSK may find its use for high-frequency communications, such as deep millimeter wave and terahertz.

Sometimes, modulation and channel coding can be designed jointly, as for trellis codes. The good design of joint coding and modulation can minimize the information loss between the modulation and channel coding when they are designed separately, and detection and decoding are conducted separately. Its performance benefit is more pronounced when SNR is high. One important criterion is the receiver complexity, which should be well contained, e.g., not too much higher than that of the traditional receiver.

For a very high SNR operating point, fast-than-Nyquist (FTN) [51] is also a good candidate and increases the spectral efficiency. However, it comes at the cost of introduction of inter-symbol interference (ISI), which needs to be suppressed or cancelled out. In this sense, FTN bears some resemblance to non-orthogonal transmission. ISI in FTN leads to an "effective" convolutional code, which operates in the complex domain. Its effective "polynomials" are determined by the waveform or modulation. Such effective convolutional code has certain error correction capability, thus reducing the reliance on the outer channel codes.

Symbol-level spreading can be used for non-orthogonal multiple access [52] so that the network can accommodate several users simultaneously. The spreading can be entirely linear, e.g., spreading followed by modulation, or joint spreading and modulation where the coded bits are mapped to the modulation constellation points directly [53]. Less advanced receivers can be used for linear spreading, for instance, minimum mean-squared error hard interference cancellation (MMSE-hard IC), where the decoder can be hard-output. In general, the cross-correlation between spreading sequences should be low in order to suppress cross-user interference. However, the lower the cross interference, the fewer are the sequences in the pool, which would increase the collision probability if spreading sequences are randomly selected. Various spreading sequences have been proposed [54]. For joint spreading and modulation, its typical receiver requires a soft-output decoder to carry out the outer iteration, as well as the inner iterations inside the multiuser detector. Its receiver complexity is, in general, noticeably higher than that of MMSE-hard IC, although its performance may be potentially better.

- **Waveform**

Since the advent of 4G, OFDM has become the prevailing waveform for mobile communication, even though it has shortcomings such as high peak to average power ratio (PAPR). Its popularity comes from 1) simple transmitter and receiver processing, to facilitate a wide use of MIMO; 2) orthogonality between sub-carriers, thus increasing the resource utilization. To alleviate the PAPR issue, DFT-s-OFDM was used in 4G and 5G to ensure the orthogonality between different users, while maintaining the single-carrier property of each user's signal, e.g., modulation symbols of each user may suffer inter-symbol interference due to the channel delay spreading. Some variants of OFDM and DFT-s-OFDM were proposed, to reduce the out-of-band emission and make signals of each user more localized in time and frequency domains. However, in 5G, all those variants became the implementation – transparent to the air interface specifications. For very high frequencies, such as Terahertz, signal digitization may become very challenging. Analog processing may find more use where a completely new waveform would be needed to make the circuit/processor feasible.

### 4.2 Visible light communications

Due to the ubiquitous LED lighting and higher than 80% penetration of the lighting market, visible light communication (VLC) based on LED has been extensively studied during the 5G research cycle. However, due to its inherent drawbacks, VLC has not been introduced to the 5G standard. However, with the in-depth investigations of next-generation lighting technologies, laser diode (LD)-phosphor conversion lighting technology with higher brightness, higher efficiency, and further illumination range than traditional LEDs is expected to be one of the most promising next-generation lighting technology [55].

It is well known that the LD can be modulated very quickly and has higher modulated bandwidth than the LED. The LD-based VLC system has been demonstrated at 28.8Gbps [56], and can potentially reach 100Gbps, which is more suitable for ultra-high data density (uHDD) services in 6G. Thus, an enhanced VLC (eVLC) based on LD lighting technologies is proposed for 6G. Moreover, the advantages of LD illumination, combined with adaptive glare-free systems, can facilitate high resolution, adaptive, programmable, and pixelated lighting system, which can basically be realized using a spatial light modulator, such as LCoS, DLP, or MEMS mirrors. These features will benefit uHDD spatial multiplexing, such as beam division multiple access, pixelated spatial multiplexing, and IDMA.

On the other hand, the beam angle of the LD is 1/5 to 1/10 of the beam angle of an LED with a similar lumen output. The small spot of the light allows the LD-based illuminator to produce sharp edges in the output, which is about ten times sharper than LED illumination. In addition, the very high directivity of LD lighting would allow longer-distance transmission between the poles compared to LED streetlights. These features may be important in some applications that are intended to illuminate only certain areas, such as outdoor applications and integrated access and backhaul (IAB).

If using a highly efficient fiber, the blue light transmittance per meter is 99.8%, so the blue light loss is very small in most lighting applications with LD-phosphor splitting and phosphor remoting. In streetlight applications, electronics and LDs can be in the road side unit (RSU) for easy maintenance, while phosphors can be placed in sealed packages on the top of the pole. In the case of enhanced VLC, flexible splits of LDs and phosphor components can achieve the convergence of fiber lighting and distributed VLC lattice for the wireless data center, resulting in a scalable, resilient, and sustainable data center network design.

In smart car applications, LD headlights offer the advantages of high luminance and compact size. As mentioned above, combining LD headlights with MEMS scanning mirror technology will be the key to enable new adaptive illumination functionality with high pixel resolution. A new concept of eVLC-based V2X is presented, which allows dynamic shaping of the basic intensity distribution of an adaptive LD-phosphor headlight. Figure 6(a) shows the V2X and V2V communications based on eVLC. In uplinks, LD headlights communicate with RSU, where eVLC receivers and LDs are embedded. In downlinks, street lamps are wirelessly connected to the eVLC receivers on top of the cars. Meanwhile, street lamps can be used to establish backhaul links with each other by free-space coherent optical communications. The eVLC-based IAB streetlamp concept is shown in Figure 6(b), in which blue laser fiber input from LDs in RSU is diverted a little for the optical integrated coherent transceiver of the backhaul link.

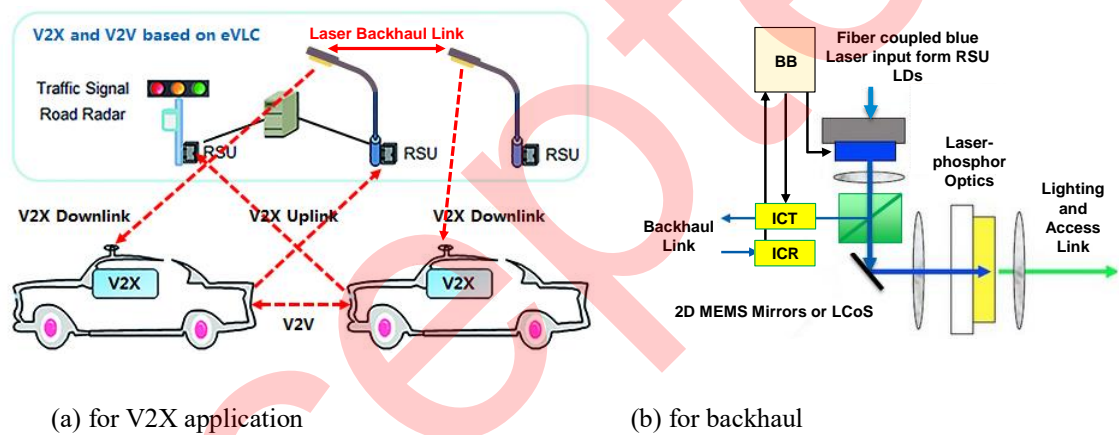

(a) for V2X application                 (b) for backhaul

**Figure 6 Two use cases for enhanced visible light communication**

### 4.3 Advanced duplex

In a practical network, one typical situation is the imbalance of the spectrum demand, between different networks, between different nodes in the same network, and between the transceiver links of the same node. These imbalances lead to the low utilization of the spectrum. Advanced duplex is expected to solve this issue. There are two potential candidate technologies: (1) spectrum sharing based (i.e., free-spectrum-sharing), mainly used to solve the imbalanced spectrum requirements between different networks, and (2) full degree of freedom duplex (i.e., free duplex), mainly for the imbalanced spectrum requirements between the transceiver links of the same node.

- **Free-spectrum-sharing**

At present, cellular networks mainly operate in licensed carriers. The owner of the spectrum resources has exclusive right to provide the access to the spectrum. Even if the spectrum resources are temporarily idle, other operators cannot use them. Such exclusive use of the spectrum imposes strict restrictions and regulatory requirements. The reason why the spectrum sharing technology has not been fully deployed is mainly attributed to the constraints of spectrum regulations, but more importantly, the lack of a matured technology for enabling spectrum-sharing. Significant breakthroughs in the research of spectrum sharing technology are needed, including efficient spectrum-sharing technology and efficient spectrum-monitoring technology, to improve the resource utilization of the spectrum in the future network by using a shared spectrum with full flexibility and to monitor the spectrum usage more conveniently. In 5G networks, AI will be adopted to improve the effectiveness of spectrum management. The 6G network is expected to be an AI-assisted autonomous system where the network resources, including the spectrum, can be used and managed more dynamically and efficiently. The combination of AI and the traditional spectrum-sharing technologies can be used to facilitate intelligent spectrum-sharing [57][58][59], i.e., free-spectrum-sharing.

- **Free duplex**

As mentioned above, because the arrival of data packets often obeys Poisson distribution, the resource utilization of transceiver links (generally referred to as uplink and downlink) in real networks fluctuates dynamically and is often unbalanced between the downlink and uplink. Enhancing the existing duplex technology means to achieve flexible spectrum allocation between transceiver links (or flexible spectrum-sharing between transceiver links), so as to increase the utilization of spectrum resources.

At present, compared to the traditional mobile communication systems, the 5G system is based on a flexible empty port concept design, while the duplex mode adopts a dynamic TDD architecture, in which the FDD mode is only a special case of configuration. In addition, 5G and B5G/6G are mainly deployed in the frequency bands above 2 GHz, most of which are TDD bands. The Cross Link Interference/Remote Interference Management Work Item Description standard project was completed in 2019 and has been included in 5G NR Rel-16 [60]. In this feature, two types of interference suppression mechanisms are introduced: a mechanism to solve the problem of cross-link interference between adjacent base stations and a mechanism to solve the problem of cross-link interference between remote base stations (cross-link interference caused by the atmospheric waveguide phenomenon). Once these two types of interference are well addressed, 5G will be able to support the commercial deployment of flexible duplex features and gradually get rid of the resource utilization constraints of fixed duplex (FDD/TDD). Although the initial technical discussion of 5G involves full-duplex technology, it has not been adopted in 5G because of its immature theoretical and technical research.

With the progress of duplex technology and its maturity in the next decade, the duplex mode in 6G era is expected to operate in the true free-duplex mode. That is, there is no FDD/TDD differentiation anymore, but a flexible and self-adaptive scheduling mode of flexible duplex or full duplex, according to the service requirements between the transceiver and transceiver links. Thus, the limitation of spectrum resource utilization between the transceiver and receiver links by duplex mechanism is lifted. The free-duplex mode can achieve a more efficient utilization of spectrum resources by sharing all-degree-of-freedom (time, frequency, and space) spectrum resources between the transceiver and receiver links (or DL and UL), so as to improve throughput and reduce transmission delay. To achieve the free-duplex mode, the key technical challenge is to break through the full-duplex technology. Figure 7 depicts the evolution of the duplex mode.

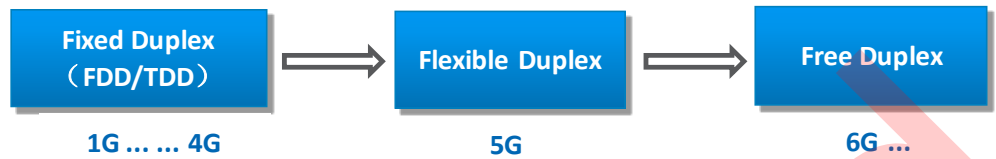

**Figure 7 Duplex evolution route of wireless mobile communication system**

Based on the technical characteristics of self-interference limitation, full-duplex technology is mainly suitable for the following typical application scenarios: (1) low-transmission-power scenarios, including short-range wireless links (e.g., device to device, vehicle to everything, and small cell with low transmission power); (2) scenarios equipped with transceiver devices without strict constraint in complexity and cost, such as wireless relay and wireless backhaul; and (3) scenarios with narrow beams and more spatial freedom, including communication scenarios using massive MIMO in the frequency bands below 6 GHz and high-frequency bands of millimeter/terahertz.

In the process of commercialization of full-duplex technology, the problems and technical challenges to be solved include suppression of high-power dynamic self-interference signal, miniaturization of self-interference suppression circuit in multi-antenna RF domain, new network architecture and interference elimination mechanism under full-duplex system, and coexistence and evolution strategy with FDD/TDD. In addition, from the perspective of engineering deployment, it is more important to study full-duplex networking technology.

## 5   Conclusion

In this paper, we first briefly discussed the conceptions of 6G, and then focused on the potential key technologies of 6G, which include four revolutionary technologies of exploratory nature and three more matured technologies. The revolutionary technologies discussed would fundamentally change the physical layer of mobile communication systems compared to 5G. Many aspects are still at the stage of scientific exploration. Yet, they represent the development level of science and technology of a country in cutting-edge strategic areas. Regarding the matured technologies, extensive study and development at the physical layer are still needed to make them feasible in engineering.

The 6G network will eventually provide terabit rate per second, support an average of 1000+ wireless nodes per person in 10 years (2030–), and provide instant holographic connectivity anytime and anywhere. The future will become a fully data-driven society in which people and things will be connected universally, almost instantaneously (milliseconds).